# From antidepressants to psychedelics: revisiting the pharmacology of so-called serotoninergic treatments

Marc Gielen*[,a], Daniela Duarte Domingues[b,c], Eric Burguière[b,c], Kevin Swierkosz-Lenart[b,d] & Luc Mallet*[,b,e,f,g]

[a] Sorbonne Université, 21 rue de l'École de Médecine, 75006 Paris, France.

[b] Sorbonne Université, Institut du Cerveau - Paris Brain Institute – ICM, Inserm U 1127, CNRS UMR 7225, AP-HP, 47 bd Hôpital de la Pitié Salpêtrière, 75013 Paris, France

[c] Lyon Neuroscience Research Center, INSERM U1028/CNRS UMR 5292/Lyon 1 University, Lyon, France

[d] Department of Psychiatry, Service of Old Age Psychiatry, Lausanne University Hospital (CHUV), University of Lausanne, Prilly, Switzerland.

[e] Laboratory of Translational Neuropsychiatry , INSERM U955 - Institut Mondor de Recherche Biomédicale (IMRB) , Créteil, France

[f] Univ Paris-Est Créteil, DMU IMPACT, Assistance Publique - Hôpitaux de Paris, Créteil, France

[g] Global Health Institute & Department of Mental Health and Psychiatry, University of Geneva, Geneva, Switzerland

* Correspondance : marc.gielen@gmail.com; luc.mallet@inserm.fr

## Abstract

Recent years have seen a resurgence of interest in serotoninergic psychedelics as potential treatment for diverse psychiatric disorders. Here, we will delve into the molecular mechanisms of drugs used to treat depression and obsessive-compulsive disorder (OCD), with implications on the mechanism of action of so-called “serotoninergic” psychedelics as well as selective serotonin reuptake inhibitors (SSRIs). In particular, we argue that the pharmacokinetic and pharmacodynamic properties of SSRIs suggest that the inhibition of the serotonin transporter (SERT) alone does not provide a comprehensive account of their clinical effects. On the contrary, the SSRI posology required to treat depression and OCD is consistent with their major molecular target(s) displaying an SSRI affinity much lower than SERT. Such is the case for the tyrosine receptor kinase B (TrkB), recently described as a putative target for drugs with antidepressant effects. TrkB might thus provide a unifying framework for understanding the therapeutic effects of antidepressant drugs. We finally discuss the pharmacological implications of this hypothesis on the therapeutical potential of psychedelics.

## Abbreviations

BDNF, brain-derived neurotrophic factor; CSTC, cortico-striato-thalamo-cortical; DMT, dimethyltryptamine; GABA, γ-aminobutyric acid; $GABA_A$ receptor, type A γ-aminobutyric acid receptor; HTR, head-twitch response; $K_d$, dissociation constant; $K_i$, inhibitory constant; LSD, Lysergic Acid Diethylamide; MDMA, 3,4-Methylenedioxymethamphetamine; NMDA, N-methyl-D-aspartate; OCD, obsessive-compulsive disorder; SERT, serotonin transporter; TrkB, tyrosine receptor kinase B; SSRI, selective serotonin reuptake inhibitor; 5HT, 5-hydroxytryptamine (serotonin)

## Data availability

No original data have been used for this Perspective. The concentration-response curves in Figure 1 were built based on data from the literature, as indicated in the text, and are available upon request.

## Acknowledgements

The authors acknowledge financial support by the Institut de Recherche en Santé Publique (IRESP) (grant IRESP-AAPSPA2021-V1-04). This research was supported by IReSP and the Institut National du Cancer (INCa) as part of a doctoral grant call for applications launched in 2022 (grant SPADOC22-008 to D.D.D.). M.G. is grateful to the Fondation Bettencourt Schueller for their support. The funders had no role in study design, data collection and analysis, decision to publish or preparation of the manuscript

## Author contributions

M.G. and L.M. conceived the manuscript and wrote the first draft. All authors contributed to the critical evaluation, further development of concepts and ideas, review and editing of the manuscript.

## Conflict of interest statement

K.S.L. regularly participates in advisory boards organized by Johnson & Johnson. The authors have no other competing interests to declare.

## ORCID

M.G.: 0000-0003-0941-4325

D.D.L.: 0009-0001-8046-1696

E.B.: 0000-0002-9690-5909

K.S.L.: 0009-0002-1809-5311

L.M.: 0000-0001-8061-1167

## Introduction

Classical psychedelics have become a major focus of research in psychiatry, with recent and on-going clinical trials spanning a large range of psychiatric disorders, from depression[1,2] to addiction[3], through anorexia nervosa[4] and obsessive compulsive disorders (OCD)[5]. Their molecular mechanism has long been depicted through their serotoninergic activity, stemming primarily from a 5HT2A receptor agonism, at least for their acute effects[6,7]. The serotoninergic properties of therapeutical drugs is specially acknowledged in OCD[8,9], and we thus decided to explore, in the literature, the pharmacology of OCD treatments as a stereotypical paradigm to question the serotoninergic contribution to their clinical effects.

OCD is a severe psychiatric condition characterized by recurring obsessive thoughts and compulsive behaviors. The lifetime prevalence accounts for approximately 2% of the general population, leading to substantial disability and a profound impact on the quality of life[10]. Current OCD treatments rely on psychotherapeutic approaches like cognitive-behavioral therapy and pharmacological interventions, typically involving high-dose selective serotonin reuptake inhibitors (SSRIs)[10,11]. Achieving symptomatic improvement through these methods is slow, usually taking several months when successful[12]. Moreover, resistance to such strategy is common, with 40% of patients showing no response to treatment[11]. Consequently, alternative modalities, such as changing antidepressants, incorporating atypical antipsychotics or glutamatergic agents, and exploring non-invasive neuromodulation techniques, are proposed, with invasive options reserved for severe cases[12]. This prompts the need to develop novel treatments.

Among the emerging substances of interest, serotoninergic psychedelics like Lysergic Acid Diethylamide (LSD) and psilocybin have garnered attention in psychiatry for their profound effects on perception, thoughts, and emotions[6]. The immediate impact of these substances on psychiatric symptoms, coupled with observed prolonged benefits weeks after a single intake, makes a compelling case for their therapeutic use[13]. From a clinical perspective, questions persist about the mode of administration and the roles of the psychotherapeutic approach versus the substance itself in symptom evolution[14,15].

Clinical evidence supporting the therapeutic efficacy of classical psychedelics in OCD remains preliminary. Case reports describe symptomatic improvement, sometimes with rapid relief lasting from days to months[16]. Retrospective naturalistic surveys also suggest benefits[17], with one reporting significant improvement with classical psychedelics but not ketamine or 3,4-Methylenedioxymethamphetamine (MDMA)[18]. Selection and recall biases nevertheless limit interpretation. An early prospective study reported rapid symptom reductions with psilocybin (25–300 µg/kg), without clear associations with dose or subjective psychedelic intensity[19]. A pharmacological-challenge study found short-term improvement, predominantly in compulsions, with 10 mg compared with 1 mg, without a parallel antidepressant effect[20]. A phase-2 randomized trial, currently available as a non-peer-reviewed preprint, found greater symptom reduction after a single 250 µg/kg dose than after niacin at one week, with improvement persisting during uncontrolled follow-up[21]. A further randomized trial of weekly psilocybin (100 or 300 µg/kg) versus lorazepam, followed by additional 300 µg/kg sessions, linked greater cumulative exposure to larger symptom reductions in a post hoc analysis[5]. These findings warrant further investigation of whether therapeutic benefit requires a minimum dose or cumulative exposure. However, small samples and limitations in blinding preclude firm conclusions about efficacy, dose–response relationships and optimal dosing.

Several mechanistic hypotheses have been proposed to explain the therapeutic potential of serotoninergic psychedelics in OCD. 5-HT2A receptor agonism may modulate maladaptive cortico-striato-thalamo-cortical (CSTC) loops implicated in compulsive behaviors[8,22], while opening a plastic or metaplastic window in which learning reshapes maladaptive habits and beliefs[8,23], with serotonergic dysfunction representing a core pathophysiological substrate[24] and dysregulated connectivity as a potential target[25]. However, the neuroimaging evidence underpinning these models remains highly speculative, and studies rarely distinguish acute pharmacological effects from sustained post-treatment changes. Critically, 5-HT2A agonism alters neurovascular coupling, decoupling hemodynamic from neuronal signals[26], challenging the validity of BOLD-based inferences during acute psychedelic states and urging caution in interpreting circuit-level conclusions. This mismatch between brief exposure and sustained benefit invites scrutiny of molecular targets supporting long-term effects.

## Pharmacokinetics and pharmacodynamics of SSRIs: old data conflicting with the 5HTT-inhibition model?

As mentioned above, SSRIs are first-line treatments for both depression and OCD, albeit at higher doses for the latter, and used as well in other disorders including anxiety and post-traumatic stress disorder. Interestingly, SSRIs' cerebrospinal fluid concentrations in the µM range are usually clinically achieved for the treatment of depressed patients[27]. For example, the blood serum concentrations measured for fluoxetine and its metabolite norfluoxetine amount to ~100 ng/mL and 120 ng/mL, respectively, in fluoxetine-treated patients[28] at doses used in treating depression. With a molar mass of ~300 $g.mol^{-1}$, this yields fluoxetine and norfluoxetine concentrations of approximately 0.3 µM achieved under usual clinical settings for depression.

How much does this translate into the brain? Based on magnetic resonance spectroscopy, early estimates of fluoxetine brain to plasma ratios ranged from 2.6 to ~20[27,29]. While those variable estimates relied on indirect measurements, a subsequent post-mortem study could directly measure, through biopsies, the distribution of fluoxetine in the corpses of ten aircraft pilots involved in deadly accidents[30]. In this study, the authors found a 15-fold enrichment of fluoxetine in the brain compared to blood.

This 2.6 to 20 brain to plasma ratio thus translates in a brain concentration in the ~0.8-6 µM range when treating depression, with a likely best estimate at around 4.7 µM based on the distribution reported in Johnson et al.[30]. Comparable concentrations have been reported for fluvoxamine when treating OCD[31] (thus with higher doses than in depression), which reaches brain concentrations in the 3-10 µM range under clinical conditions. However, the SERT affinity for SSRIs is in the nanomolar range, with a $K_i$ (inhibitory constant) ranging from 0.7 nM for paroxetine to 6.2 nM and 14 nM for fluvoxamine and fluoxetine, respectively[27]. In other words, clinically achieved brain concentrations of SSRIs exceed the affinity of these compounds for SERT by approximately two orders of magnitude.

If the serotonin transporter (SERT) was the main clinically relevant target for SSRIs' effects in depression, clinically achieved concentrations would be largely sufficient to fully saturate the SERT SSRI binding site at a fraction of antidepressant doses (Figure 1A). Thus, an antidepressant effect of SSRIs should be attained at much lower doses than the ones clinically necessary: a tenth of a fluoxetine pill (i.e. 2 mg) taken daily should be enough to lead to a SERT occupancy of approximately 85%-95% (Figure 1A), and thus to achieve near maximal clinical response. This is not what is observed clinically[32]. Moreover, further increasing the SSRIs dose would not be needed to improve treatment of OCD (Figure 1A). The requirement of SSRI doses used in the treatment of depression, as well as the recommendation for higher SSRI doses in OCD, thus likely involve a molecular target with a lower affinity for SSRIs than SERT, probably in the micromolar range (i.e. ~100-fold lower affinity than that of SERT).

Along the same lines, Meyer et al. reported more than twenty years ago that a daily dose of 2 mg of fluoxetine for four weeks lead to an occupancy of SERT reaching half of its maximal effect in a positron emission tomography (PET) study using [$^{11}$C]N,N-Dimethyl-2-(2-amino-4-cyanophenylthio) benzylamine ([$^{11}$C]DASB)[33]. This may appear significantly different from our 85%-95% estimated range. However, the interpretation of a reduced [$^{11}$C]DASB PET signal by SSRIs is not straightforward: it possibly reflects both a competition between fluoxetine and [$^{11}$C]DASB and a downregulation of SERT following the four-weeks long daily treatment, set aside any potential non-specific binding of [$^{11}$C]DASB. Also, maximal SERT occupancy by SSRIs reached ~85% in this study and not 100%. If [$^{11}$C]DASB is truly specific for SERT, this means that there is no direct competition between SSRIs and [$^{11}$C]DASB, but rather a negative allosteric interaction, complicating the analysis of the dose-response of SSRI competition with [$^{11}$C]DASB at SERT. Still, the strong occupancy of SERT by SSRIs with subclinical doses, and the fact that this occupancy reaches its maximum below the minimal recommended therapeutical doses, further highlights that antidepressant effects of fluoxetine are only achieved above doses fully saturating SERT. Moreover, with these data in mind, one should not expect any supplemental clinical effect of increasing the doses above 20 mg fluoxetine, which contradicts the requirements for higher doses in OCD. Finally, Meyer at al. strikingly reported in their PET study that, at SSRI-treated depressed patients, there was no relationship between the SSRI occupancy of SERT and changes in Hamilton depression scale[33]. If one assumes that the clinical effects of SSRIs stem from their serotonin-reuptake properties, we expect on the contrary a correlation between SERT occupancy by SSRIs and therapeutical response.

One should note that SERT itself harbors a second, low-affinity site in its extracellular vestibule, occupied in the micromolar range, where a second inhibitor molecule stabilizes the drug bound at the high-affinity central site and slows its dissociation[34,35]. In other words, the occupancy of the low-affinity site further increases the affinity for the high-affinity site. Such mechanism does not increase the extent of transport blockade — already complete once the high-affinity site is saturated — and therefore offers no straightforward account of why higher SSRI doses should confer greater clinical benefit.

## Direct binding to TrkB as a unified action mechanism for antidepressants?

Interestingly, a recent study proposes a unified pharmacological target for antidepressants. Indeed, Casarotto et al. showed in 2021 that TrkB, the main neuronal receptor for the brain-derived neurotrophic factor (BDNF), binds a series of antidepressants, including typical antidepressants like SSRIs (fluoxetine) and tricyclic antidepressants (imipramine) as well as the fast-acting antidepressant ketamine, also considered a non-classical psychedelics and largely viewed as an NMDA receptor antagonist[36].

This hypothesis is particularly compelling given that BDNF and TrkB form a major signaling system regulating neuronal development, neuritogenesis and synaptic plasticity[37] involving downstream mammalian target of rapamycin (mTOR) activation[38]. Moreover, the therapeutical effects of psychedelics have been suggested to involve neuroplasticity[39,40]. A recent preclinical study performed in rats also showed that a single dose of LSD enabled a stronger and longer-lasting change in brain activity following direct infralimbic cortex stimulation, together with a higher mTOR activation, as expected for increased plasticity[41].

In addition to computational and biophysical experiments to study the TrkB-antidepressant interaction, the authors examined the behavioral effects of antidepressants on both wild-type mice and knock-in mice (TrkB-Y433F), in which the TrkB antidepressant binding sites were mutated. Using several experimental paradigms, they found that the antidepressant effects of fluoxetine were abolished in the knock-in mice, suggesting that TrkB, rather than SERT, may be the more relevant target of SSRIs. The effects of ketamine, assessed with the fear conditioning paradigm, were also lost in the knock-in mice, also raising the possibility that the main therapeutical target for ketamine in the brain might be TrkB rather than NMDA receptors.

It should however be noted that the TrkB-Y433F mutation not only impairs the binding of antidepressants to TrkB, but also slightly affect some other biophysical properties of TrkB[36]. As such, the results obtained with knock-in animals should be interpreted with caution. Indeed, if TrkB-Y433F has a broader impact on BDNF-induced TrkB signaling, it remains possible that such mutation indirectly disrupts the effects of antidepressants on downstream BDNF signaling. In a subsequent study, the authors provide evidence that the TrkB-Y433F mutation does not lead to global TrkB activation, nor does it lead to behavioral effects in the absence of treatment by antidepressants[42], suggesting that baseline BDNF signaling is unaffected by the TrkB-Y433F mutation. Still, further work and reproduction by other groups will be critical to determine the relative weight of direct TrkB binding *versus* binding to other "classical" targets (e.g., SERT for SSRIs, NMDA receptors for ketamine) in the clinical effects of antidepressants.

Strikingly, the TrkB affinity for SSRI is in the micromolar range[36], and therefore more sensitive to SSRIs concentration variations around the clinically achieved concentrations when treating depression and OCD (Figure 1A), as stated above. One is thus tempted to speculate that TrkB is a key driver for the efficacy of SSRIs in OCD. In such hypothesis, ligands with higher affinities for TrkB would hold promising therapeutical potential value for the treatment of OCD.

## TrkB and the pharmacology of classical psychedelics: relevance to the treatment of depression and OCD?

Subsequently, the same research group expanded this work on TrkB with the study of psychedelics, specifically psilocin (the active metabolite of psilocybin) and LSD[43]. These compounds are thought to primarily target 5HT2A receptors, while they actually modulate numerous receptors (see Box 1). Moliner et al. showed that i) psilocin and LSD directly bind to TrkB; ii) the binding of LSD and psilocin is reduced by almost two orders of magnitude at the TrkB-Y433F mutant; iii) the psychedelic-induced neuroplasticity at murine neuronal cultures is prevented by the TrkB-Y433F mutation, as assessed in neuronal cultures from TrkB-Y433F$^{+/-}$ heterozygous mice; and iv) TrkB plays a key role in mediating the anti-depressant properties of LSD in mice, as assessed in the forced swim test and in fear conditioning experiments, with a behavioral effect at wild-type animals that is lost at TrkB-Y433F$^{+/-}$ heterozygous mice. Of course, the same limitations apply as for the previous study on antidepressants, namely a potential indirect effect through modifications of the TrkB signaling pathway due to the TrkB-Y433F mutation. We should emphasize here that these data were drawn from preclinical murine model, keeping in mind questions regarding the robustness and the validity of these models (Box 2).

Interestingly, the affinities of psilocin and LSD for TrkB are three orders of magnitude higher than SSRIs[36,43]. One could thus speculate a therapeutic potential for psychedelics in the treatment not only of depression, but of OCD as well. LSD indeed binds TrKB in the low nanomolar range in direct radioligand-binding experiments (~0.93 nM[43]), an affinity very similar to the one reported for 5HT2A receptors using a similar direct binding measurement of radiolabeled LSD, estimated at 0.33 nM[44] (Figure 1B). However, it is particularly important to stress that different experimental paradigms lead to different "affinity" values. As such, one should only compare affinities measured using identical experimental settings and avoid direct comparisons between dissociation constant (Kd) and inhibitory constant (Ki) measurements (Box 3).

It is interesting to note that, in the study by Moliner and colleagues[43], both Kd and Ki measurements have been performed for LSD at TrkB: direct binding measurements of radiolabeled LSD result in a Kd value of 0.93 nM (Figure 1B), while Ki measurements (defined in their study as displacement of radiolabeled-LSD by non-labeled LSD) provide a Ki of 3.4 nM (Figure 1C). This latter value is probably a better proxy of LSD binding properties when comparing with studies performed on 5HT2A receptors using Ki measurements.

Of note, psilocin direct binding experiments have not been performed on TrkB in the Moliner study[43], and not even on 5HT2A receptors to our knowledge; only Ki measurements are currently available. Those lead to a psilocin Ki value of ~6.7 nM at TrkB, similar to the 3.4 nM Ki value reported for LSD, using the same experimental approach[43], and should not be directly compared with the apparently higher LSD affinity observed in direct binding experiments.

When comparing psilocin Ki values between TrkB and human 5HT2A receptors, the latter seem to display an apparent affinity almost an order of magnitude lower, at about 50 nM[45] (Figure 1C). Given that psilocin occupies up to 72% of human brain 5HT2A receptors at clinically relevant doses[46], it can be inferred that TrkB psilocin sites' saturation is almost achieved under those conditions, making it a promising candidate for OCD in the hypothesis that the primary target for SSRIs is indeed TrkB.

We should however exert caution with possibly oversimplified schemes for the various clinical effects of psychedelics, and the TrkB hypothesis for psychedelics' actions has yet to be reproduced by other groups – there is bound to be some controversy regarding the main molecular targets for the antidepressant effects of hallucinogenic psychedelics. A recent paper by the Kwan lab[47] indeed showed that psilocin induces structural plasticity in mice frontal cortical pyramidal neurons *in vivo*, and that such plasticity is eliminated through targeted knock-out of 5HT2A receptors in pyramidal tract neurons. Moreover, they showed that knocking-out 5HT2A receptors in the medial frontal cortex of mice eliminates psilocybin-induced anxiolytic effects assessed in the tail-suspension test. It should however be noted that the tail-suspension test was, in this study, performed one day after psilocybin administration, in stark contrast with the Moliner et *al*. paper where the effect of LSD on the forced swimming test was evaluated 7 days after administration, and where the LSD-induced extinction of conditioned fear response was evaluated 4 weeks after administration. Finally, in Shao et *al*.[47], the effects of psilocybin on learned helplessness (also evaluated one day after psilocybin administration) did not reach statistical significance, neither in control animals nor in local 5HT2AR knock-out, precluding any interpretation on this trait.

To reconcile both views, one could speculate that the full clinical efficacy of antidepressants relies both on TrkB as a primary molecular target, and on the binding to other receptors as secondary targets favoring BDNF signaling, thereby interacting in synergy with TrkB direct binding: i) 5HT2A receptors in the case of psychedelics, which in turn increases BDNF signaling[39]; ii) SERT in the case of SSRIs, which will also increase BDNF signaling through, potentially, serotonin-induced 5HT2A activation[48]; or iii) NMDA receptors in the case of ketamine, which will also increase BDNF signaling[49]. Such scheme would be consistent with a recent paper showing that serotoninergic psychedelics promote dendritogenesis, synaptogenesis and c-Fos and Egr-2 gene expression that require both TrkB and 5HT2A receptors[50]. In the case of SSRIs, if one assumes that both SERT inhibition and TrkB modulation are necessary for therapeutic efficacy, the required posology for SSRI administration would be dictated by the target with the lowest affinity, namely TrkB.

The binding curves displayed in Figure 1C also raise some questions regarding psychedelics microdosing. For LSD, the binding affinities are almost identical on TrkB and 5HT2ARs (Figure 1B&C). In the hypothesis that the main driver of the long-term antidepressant effects of LSD is mediated by TrkB, a microdosing that doesn't elicit a psychedelic experience should not lead to a significant antidepressant effect. This discussion is potentially more interesting in the case of psilocybin: for its active metabolite psilocin, the affinity for TrkB is reported to be approximately an order of magnitude higher than for 5HT2ARs (Figure 1C). Thus, in theory, it could be possible to significantly increase the TrkB signaling while 5HT2ARs would be mostly unoccupied (for example within the 1-3 nM range for psilocin binding to TrkB and 5HT2ARs in Figure 1C).

However, regarding psilocybin microdosing, it is important to note that the binding affinity of psilocin is more than an order of magnitude higher for 5HT2B than for 5HT2A receptors[51], with a psilocin Ki value of ~5 nM for 5HT2B receptors[52]. Given that 5HT2B agonism has been involved in the development of cardiac fibrosis and valvulopathy, this raises the possibility that repeated psilocybin usage, as seen in microdosing protocols, might be dangerous from a cardiac point of view[52].

## Future directions

To foster the debate on the clinically relevant targets for serotoninergic psychedelics, one could use the 5HT2AR antagonist ketanserin in clinical trials. Indeed, ketanserin is FDA approved and prevents the hallucinogenic effects of serotoninergic psychedelics[53]. An obvious and interesting clinical trial would thus be to assess whether serotoninergic psychedelics retain their long-lasting antidepressant effects when preventing the hallucinogenic experience with ketanserin. Interestingly, a current clinical trial (https://euclinicaltrials.eu/ctis-public/view/2024-512911-34-00?lang=en) is examining the antidepressant efficacy of a single-dose of dual administration of psilocybin and trazodone, the latter drug being an antagonist of 5HT2A receptors.

While, in this protocol, a single dose of trazodone is used to acutely block 5HT2A receptors and examine the antidepressant properties of psilocybin when inhibiting the psychedelic experience, it is striking that trazodone is usually used as an antidepressant when administered daily. On top of its 5HT2A antagonist properties, it also inhibits SERT, among other targets. However, its apparent affinity for 5HT2ARs ($K_d$ ~ 7.7 nM) is actually about twentyfold higher than for SERT (Kd ~ 189 nM), and its affinity for SERT is more than tenfold lower than that of fluoxetine[54]. These observations potentially hint towards a role of the 5HT2A antagonism in trazodone's antidepressant properties, although it is unclear what are the trazodone brain concentrations achieved clinically. Such mechanism may appear contradictory with an antidepressant role of classical psychedelics through their 5HT2A agonism.

Given that psychedelics lead to antidepressant effects at patients experiencing a hallucinogenic experience, it is often concluded that the antidepressant efficacy requires a psychedelic experience, or that antidepressant activity and psychedelic experiences share common molecular signaling pathways[7]. Such interpretation equates to framing a correlation as a causal link. It only says that the molecular target for the antidepressant activity of psychedelics displays an affinity for psychedelics similar to the target for the hallucinogenic activity. As such, the similar affinity of psychedelics for 5HT2A and TrkB could be consistent with such observation – although it doesn't prove by any mean that TrkB is the definitive target for the antidepressant properties of psychedelics.

One could argue about the utility of a non-hallucinogenic psychedelic-derived treatment for depression or OCD. Three points are relevant to us in this regard.

First, current trials of psychedelics pose a methodological issue: blinded trials are basically impossible, as participants can clearly report whether they have been subjected to psychedelic experience[2,7]. Retaining antidepressant (and anti-OCD?) properties without the need for a psychedelic experience would greatly help in the design of double-blind randomized trials. On this topic, it is unclear to us why a single high dose of ketamine hasn't been used so far as a placebo in serotoninergic psychedelics trials: sustained responses require repeated administrations and maintenance treatments, and there is even some debate regarding the efficacy of ketamine as an antidepressant[55,56].

Second, it is unclear to which extent the psychedelic experience would prevent the scalability of psychedelic use in clinical settings. Indeed, we do not know what fraction of patients would be reluctant to undergo a psychedelic experience. This might explain why, in a recent clinical trial assessing the efficacy of psilocybin on treatment-resistant depression in a public healthcare setting, participants were highly educated[2], thus suggesting a correlation between education level and psychedelics acceptance. Moreover, for those who would accept it, such treatment would require the continuous presence of a therapist for the duration of the psychedelic experience, i.e. for several hours, in healthcare setting[2]. Such labor-intensive procedure may complicate the use of psychedelics in a large scale, which is currently a matter of intense debate[7]. As such, psychedelic-derived compounds devoid of hallucinogenic properties might be easier to deal with, like the recently described 2-bromo-LSD[57].

Third, the psychedelic experience may put at risk patients with a psychotic disorder, possibly precluding the therapeutic use of LSD or psilocin in depressed patients with schizophrenia or with an at-risk mental state. It should however be noted that the risk of psychedelics use for patients with schizophrenia is, again, a debated topic: a recent meta-analysis suggests that it may have been overstated, and that current evidence does not enable its correct assessment[58]. Still, a more conservative approach with a non-hallucinogenic psychedelic-derived drug might be safer to use for these patients. Such strategy could be particularly relevant, given the reported long-lasting effects of psychedelics in treating depression. Taking a medication daily, as currently done with SSRIs, might prove difficult for patients with a high level of disorganization. On the contrary, a medication that must be administered once a month or even less frequently could warrant a much higher adherence.

## Box 1: The multiplicity of pharmacological targets of classical psychedelics

We want here to stress the multiplicity of psychedelics' pharmacological targets. One should thus be wary of attempts to circumscribe their clinical effects to a single biological pathway.

The usually portrayed molecular targets for classical psychedelics and nonclassical ones like ketamine and MDMA have been extensively reviewed in a recent article[59], including the pharmacological specificities observed for the various classical psychedelics. Our aim will be here to briefly focus on a few main known targets for psychedelics, highlight recent developments and question potential targets for psychedelics that still need to be investigated.

### 5HT2A receptors

For the past few decades, psychedelics' actions have been depicted as resulting from their direct agonist properties at the 5HT2A receptors. This classical mechanism has been extensively reviewed elsewhere[6,59], and will not be addressed in detail hereby. Given the expression levels of 5HT2A receptors in most cortical and sub-cortical regions[60], such receptors are clearly involved in the physiopathology of the cortico-striatal-thalamic-cortical (CSCT) loop. Moreover, 5HT2A receptors are also expressed on dopaminergic neurons in the VTA, thereby promoting dopamine release in the prefrontal cortex through the mesocortical pathway and in the ventral striatum through the mesolimbic pathway[61].

Some studies suggest that 5HT2A receptors are one of the clinically relevant targets of selective serotonin-reuptake inhibitors (SSRIs) in the treatment of depression. Indeed, depressed patients carrying a Single-Nucleotide Polymorphism on the 5HT2A gene show a diminished response to SSRI treatment. Moreover, 5HT2A knock-out mice are resistant to SSRI treatment in animal models for depression, including novelty suppressed feeding tests and tail suspension tests[62]. It is thus tempting to hypothesize that those receptors are also involved in the therapeutical actions of SSRIs in the treatment of OCD. In such scheme, psychedelics would also prove useful in the treatment of OCD.

Of note, psychedelics are also active at 5HT2B and 5HT2C receptors[59,63]. It has been suggested that 5HT2C receptors are involved in the effects produced by psychedelics, especially at high doses[64]. On the other hand, binding to 5HT2B receptors rather raises concerns regarding the cardiac safety of repeated uses of psychedelics (see Main Text).

### Intracellular pools of 5HT2A receptors

While the 5HT2A-centric view of psychedelics has focused on the synaptic signaling driven by receptors expressed at plasma membranes, a recent study suggests that psychedelics actually exert their antidepressant effects through intracellular pools of 5HT2A receptors found mostly at the Golgi apparatus[65]. In such hypothesis, psychedelics owe their effects to their hydrophobicity, which allows them to cross the plasma membrane, whereas the hydrophilic properties of serotonin would not allow for psychedelics-like experiences. The activation of this intracellular pool of 5HT2A receptors by psychedelics, including psilocin and DMT, was then shown in vitro to promote an increase in the dendritic arbor complexity in cultured rat embryonic cortical neurons[65], arguing in favor of a 5HT2A-driven neuroplasticity induced by psychedelics in cortical regions. In a mouse model where the authors expressed SERT in the mPFC to enable serotonin to cross the plasma membrane of cortical neurons, promoting serotonin release with PCA ((±)-para-chloroamphetamine) led to antidepressant effects, seemingly because serotonin could thus reach intracellular pools of 5HT2A receptors[65]. Interestingly, correlations between depression and serotonin metabolites concentration in the cerebrospinal fluid have been classically interpreted as suggesting that extracellular serotonin is a key driver of depression[66]. These observations could suggest that intracellular concentrations of serotonin are actually more relevant.

## 5HT1 receptors

Psychedelics also bind to other 5HT receptors: LSD is a 5HT1A receptor agonist, explaining why acute LSD administration decreases the firing of 5HT neurons from the dorsal raphe nucleus (DRN), via 5HT1A receptors' activation[61]. Similarly, psilocin binds to 5HT1A receptors, and induces the same firing decrease on 5HT/DRN neurons. It is also striking that repeated administration of low-dose LSD in mice increases the firing of 5HT/DRN neurons, presumably through the persistent activation and subsequent desensitization of 5HT1A receptors[67]. At the behavioral level, the same protocol of repeated low-dose LSD for 7 days prevents stress-induced anxiety in mice. Such mechanism is highly reminiscent of the hypothesis of 5HT1A receptors' desensitization in the antidepressant actions of SSRI.

Indeed, in depression, the increased extracellular concentration of serotonin following chronic use of SSRI has been suggested to involve a major contribution from 5HT1A receptors' desensitization. In a bipartite model of brain serotonin function, Carhart-Harris & Nutt argue that 5HT1A receptors are the main target of SSRI treatment and enhance the tolerance to stress, while 5HT2A receptors' activation following psychedelics intake would mostly enhance the adaptability to change[68].

However, case reports and clinical trials examining the effectiveness of LSD and psilocybin in depression are usually designed with a single acute administration of psychedelics, potentially repeated once after a few days or weeks. The promising results obtained so far are thus unlikely to involve 5HT1A receptors' long-term desensitization.

## Other metabotropic receptors

In addition to their effects on serotoninergic signaling, psychedelics modulate other metabotropic receptors. In particular, LSD has been shown to exert pharmacological effects on different targets in a dose-dependent manner: while low doses mostly stimulate 5HT2A and 5HT1 receptors, higher doses also stimulate dopamine D2 receptors, as well as trace amine-associated receptor 1 (TAAR1) receptors, another class of metabotropic receptors involved in the modulation of both serotoninergic and dopaminergic systems[69]. Another type of metabotropic signaling is also capable of providing a cross-talk between serotoninergic and dopaminergic systems: indeed, dopaminergic D2 and serotoninergic 5HT2A receptors can form heteromeric complexes, thereby enhancing the pharmacological repertoire of psychedelics' targets[59]. However, such heteromeric metabotropic signaling is more likely to play a role in pharmacological differences amongst classical psychedelics rather being a shared mechanism for psychedelics' clinical effects: unlike LSD, other psychedelics like psilocybin do not bind D2 dopaminergic receptors[59].

Until recently, the existence of heteromeric complexes comprising both mGluR2 metabotropic receptors and 5HT2A receptors was also put forward, to explain the inhibition of psychedelics signaling by the activation of mGluR2 receptors[70]. However, there is a current controversy around this topic, as a recent study (in preprint) recently found no evidence for direct physical interaction, colocalization or oligomerization of the two receptors, nor any evidence that mGluR2 receptors alter 5HT2AR ligand binding[71]. The authors of this latter study hypothesize a functional inhibition of 5HT2A-containing pyramidal neurons, through presynaptic inhibition of glutamate release by mGluR2 receptors.

## Sigma-1 receptors and implications for neuronal communication

Some psychedelics like dimethyltryptamine (DMT), ayahuasca-derived compounds and ibogaine have been shown to activate Sigma-1 receptors[59,72]. This elusive type of receptors is found widely expressed in the dorsal striatum, in the orbito-frontal and anterior cingular cortex and in the thalamus, i.e. the entire cortico-striatal-thalamo-cortical (CSTC) loop structures that are critical in the pathophysiology of OCD. These intracellular receptors, expressed in MAMs (Mitochondria-Associated ER Membranes), act as ligand-modulated molecular chaperones[73,74]. Nuclear Sigma-1 receptors also have a role in transcription regulation by recruiting chromatin-remodeling factors[74]. As such, their activation has multiple effects in calcium-homeostasis[73], the expression of voltage-gate ion channels[75], and the expression of key ligand-gated ion channels such as the glutamate-gated NMDA receptors[76]. Of note, Sigma-1 receptors can be activated both by psychedelics and neurosteroids[77] , hinting potentially towards similarities between the binding sites of these compounds on Sigma-1. Given the modulation of NMDA and GABAA receptors by neurosteroids[78,79], this also raises the possibility that psychedelics could directly modulate the excitation/inhibition neuronal balance through direct binding to NMDA or GABAA receptors. This contrasts with the lack of data examining the potential modulation of NMDA or GABAA receptors by serotoninergic psychedelics in recombinant systems in the literature to our knowledge, and we feel such direct modulation should be investigated.

## Box 2: Animal models

Animal models provide an essential, but necessarily partial, bridge between clinical observations and mechanistic hypotheses. In mice, acute psychedelic administration induces the head-twitch response (HTR), a rapid side-to-side head movement primarily mediated by 5-HT2A receptors (5-HT2ARs)[80,81], and HTR frequency correlates with hallucinogenic potency in humans[82]. It should be stressed, however, that many non-hallucinogenic drugs do elicit HTRs, like, for example, benzodiazepines and atropine[83]. Moreover, this assay captures only one pharmacological dimension of psychedelics. Species differences in 5-HT2AR binding — including an Ala242 residue in mice versus Ser242 in humans leading to faster LSD dissociation in rodents[84] — limit direct extrapolation from rodent dosing schedules to long-lasting human clinical effects. Recent preclinical strategies have begun to address these limitations by testing repeated dosing regimens[67,85], developing psychedelic-inspired biosensors[86] and engineering humanized 5-HT2AR mice[87]. At the same time, long-term psychedelic effects may also involve signaling through receptors other than 5-HT2ARs, including 5-HT2C receptors in mice[88] and humans[89], or tyrosine receptor kinase B (TrkB) receptors[43] (see Box 1 for other molecular targets).

Interestingly, Moliner et *al.* reported that the psychedelics-induced head-twitch response (HTR) is unaffected by the TrkB-Y433F mutation used to impair antidepressant responses, while it is fully inhibited by M100907, which is a 5HT2A receptor antagonist. On the contrary, M100907 treatment does not seem to inhibit the psychedelics-induced neuroplasticity on neuronal cultures, nor does it inhibit the LSD antidepressant properties in behavioral experiments. Given that HTR is a widely accepted mouse proxy for the hallucinogenic effects of psychedelics, such data clearly raises the possibility to separate the hallucinogenic effects of psychedelics from their antidepressant effects, the latter putatively resulting from TrkB binding while the former are probably due to their 5HT2A agonism. This could hint towards the design of psychedelics-based antidepressants devoid of hallucinogenic effects. In that respect, it would be of interest to examine the TrkB binding of the recently developed 2-Br-LSD. This LSD analog shows impaired 5HT2A agonism compared to LSD, and does not induce head-twitch response at wild-type mice while retaining antidepressant properties assessed in forced swim test experiments[57]. However, one should note that the behavioral antidepressant effects of 2-Br-LSD are inhibited by the 5HT2A antagonist M100907, suggesting that 5-HT2A signaling may still contribute to its therapeutic-like actions.

Translational interpretation is further complicated in OCD animal models because repetitive behaviors are not pathological by default: self-grooming and digging are species-typical, goal-directed behaviors involved in hygiene, thermoregulation, foraging or shelter building[90]. Their relevance to compulsivity depends on excessiveness, rigidity, loss of adaptive function and persistence despite adverse consequences[91,92]. A full discussion of how repeated behaviors should be operationalized and interpreted in rodents is beyond the scope of this paper. Here, we simply note that OCD-like phenotypes are most often assessed through the compulsive dimension, using excessive grooming or digging in genetic models such as SAPAP3- or SLITRK1-deficient mice, or in pharmacological models such as quinpirole or 8-OHDPAT exposure[93]. In this context, SAPAP3-KO mice remain particularly informative because they combine compulsive-like over-grooming with cortico-striatal abnormalities relevant to OCD[94]. Two independent laboratories recently reported enduring reductions in compulsive-like grooming after a single dose of psilocybin in SAPAP3-KO mice[95,96]. However, these studies differed in dose, compound formulation, time window, sex effects and control design, cautioning against treating reduced grooming as a unitary or disease-specific readout. Thus, preclinical work is indispensable for dissecting psychedelic mechanisms, but animal behaviors should be used as mechanistic probes rather than direct proxies for human OCD symptoms.

## Box 3: “affinity” measurements under the lens of allostery

When comparing binding affinities from different sets of experiments, it is important to consider differences in experimental settings. Most studies examining the binding of psychedelics to 5HT2A receptors rely on inhibitory constant (Ki) measurements, whereby the studied psychedelics’ binding properties are estimated by their capability to displace another radiolabeled ligand, whether a competitive antagonist such as ketanserin or another agonist such as LSD. Those are different measurements compared to the measurement of direct binding constant (Kd) where the studied ligand is radiolabeled itself. Both approaches integrate not only binding properties but also the effects of receptors' conformational changes. Indeed, an agonist such as LSD promotes the receptors' activation, i.e. favors the occupation of activated states, which display a higher binding affinity for agonists than resting states, as accounted for in the allostery framework[97]. The "binding" affinity of ligands in direct binding experiments thus measures both the contribution of binding properties, and the capability of the ligand to promote conformational changes. For a binding experiment to properly measure pure binding properties of a ligand to a receptor, it is thus important to examine a ligand that does not promote any conformational change. That is the case for competitive antagonists, but not for agonists[98]. This is even more complicated in displacement experiments, where the displaced labeled ligand is itself an agonist, i.e. a ligand that promotes entry into high-affinity active states. In other words, for agonists or allosteric ligands, no direct binding "experiment" measures true affinity values for a given state of a receptor.

Of note, the above-defined terminology for Ki vs Kd at 5HT2A receptors is different from the usual denomination of Ki for enzymes and transporters. For transporters such as SERT, Kd are often drawn from competition assays between the studied ligand and a radio-labeled competing ligand[99], while Ki values examine the relationship between the studied inhibitor (e.g. the SSRI) concentration and the functional inhibition of the radiolabeled transported substrate (i.e. serotonin for SERT)[27]. Reported values in the literature, for fluoxetine activity at SERT, actually point towards a 17-fold higher "affinity" when measuring Kd (0.81 nM[99]) compared to Ki (14 nM[27]). It is thus of paramount importance to compare "affinities" measured with similar experimental approaches.

## Figure legends

**Figure 1: Simulated concentration-response curves for fluoxetine, psilocin and LSD to their canonical targets and TrkB. A**, simulated concentration-response curves for fluoxetine binding to the serotonin transporter SERT (blue) and TrkB (red), using Hill equations with a Hill coefficient of 1 and binding affinities of 14 nM for SERT (based on the inhibitory constant Ki[27]) and 2.4 µM for TrkB (based on direct binding of biotinylated fluoxetine[36]). The dotted box represents the approximate range of fluoxetine concentrations clinically achieved in the brain when treating depression (see Main Text), while the grey box is an approximation of the ranges achieved in treating OCD, assuming a three-fold increase in the fluoxetine concentrations in the brain compared to the posology used in depression. It is important to stress that the use of Ki values provides a lower fluoxetine affinity for SERT than direct binding experiments, with a Kd of 0.81nM[99] (see also Box 3). The use of Ki data in the figure for SERT thus might underestimate the true relative difference between fluoxetine binding properties at SERT vs TrkB. The dark orange range displays the expected range for brain fluoxetine concentrations achieved using a daily dose of 2 mg fluoxetine, assuming a 10-fold decrease compared to the range attained in depression treatment (usually with 20 mg fluoxetine[28]). **B**, simulated concentration-response curves for radiolabeled LSD binding to 5HT2A receptors (orange) and TrkB (black), using Hill equations with a Hill coefficient of 1 and binding affinities of 0.33 nM for 5HT2A[44] and 0.93 nM for TrkB[43]. **C**, simulated concentration-response curves for LSD and psilocin binding to 5HT2A receptors and TrkB based on inhibitory constant (Ki) measurements. Hill equations with a Hill coefficient of 1 were used, as well as LSD binding affinities of 3.5 nM for 5HT2A receptors[100] (orange) and 3.4 nM for TrkB[43] (black) and psilocin binding affinities of 50 nM for 5HT2A[45] receptors (green) and 6.7 nM for TrkB[43] (grey).

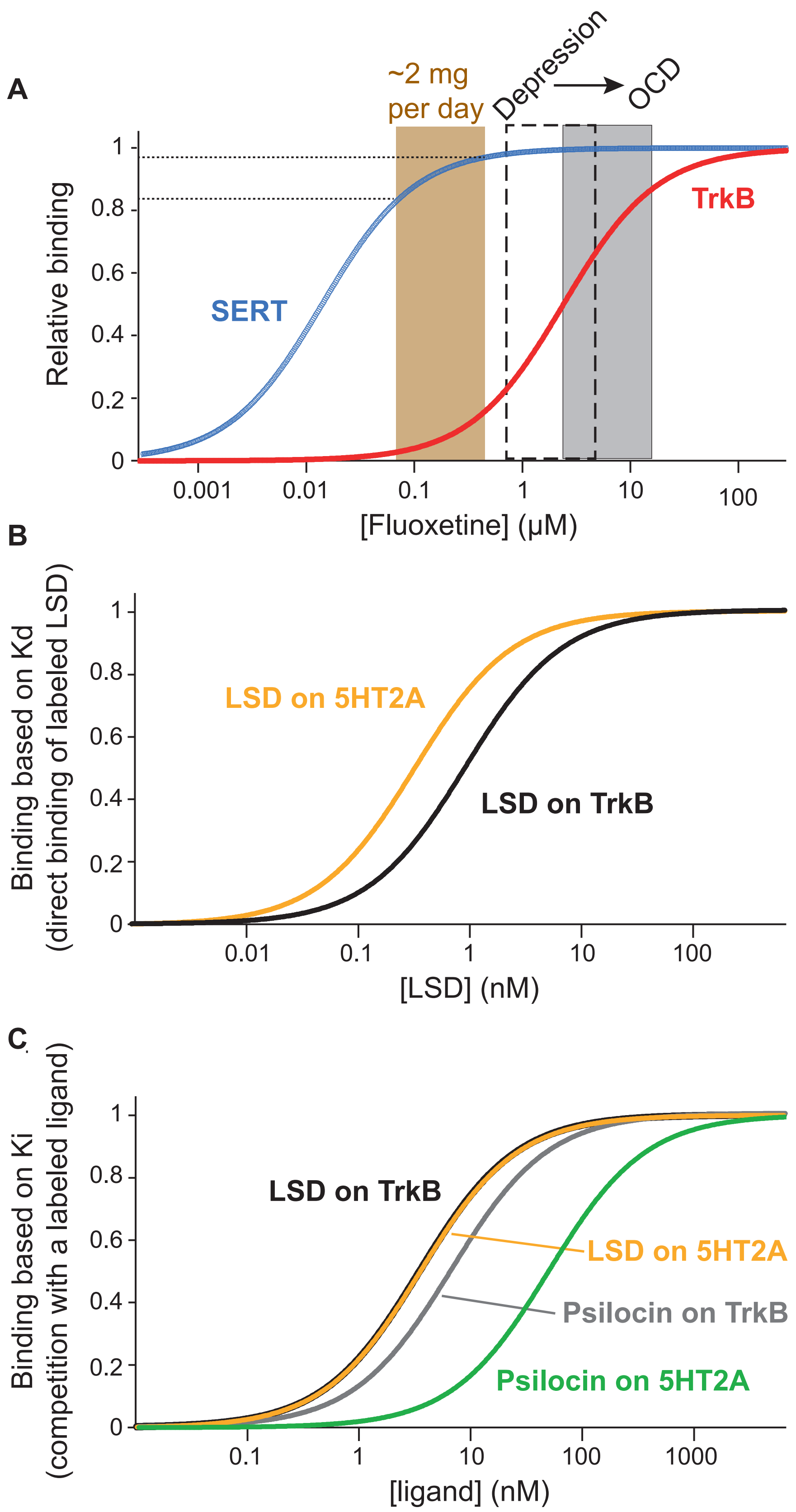
A
~2 mg
per day
Depression
OCD
Relative binding
1
0.8
0.6
0.4
0.2
0
SERT
TrkB
0.001
0.01
0.1
1
10
100
[Fluoxetine] (μM)
B
Binding based on Kd
(direct binding of labeled LSD)
LSD on 5HT2A
LSD on TrkB
0.01
0.1
1
10
100
[LSD] (nM)
C
Binding based on Ki
(competition with a labeled ligand)
LSD on TrkB
LSD on 5HT2A
Psilocin on TrkB
Psilocin on 5HT2A
0.1
1
10
100
1000
[ligand] (nM)

**Figure 1**